\documentstyle[12pt,epsfig,rotate]{article}
\input{psfig}

\newcommand{\gsim}{\;\rlap{\lower 3.5 pt \hbox{$\mathchar \sim$}} \raise
1pt \hbox {$>$}\;}
\newcommand{\lsim}{\;\rlap{\lower 3.5 pt \hbox{$\mathchar \sim$}} \raise
1pt \hbox {$<$}\;}

\newcommand{\bea}{\begin{eqnarray}}
\newcommand{\eea}{\end{eqnarray}}
\newcommand{\bd}{\begin{displaymath}}
\newcommand{\ed}{\end{displaymath}}
\newcommand{\be}{\begin{equation}}
\newcommand{\ee}{\end{equation}}
\newcommand{\ord}{{\cal O}}

\newcommand{\mt}{m_{\rm t}}
\newcommand{\mc}{m_{\rm c}}

\newcommand{\mw}{M_{\rm W}}
\newcommand{\gev}{\, {\rm GeV}}
\newcommand{\mev}{\, {\rm MeV}}

\newcommand{\vcb}{|V_{cb}|}

\newcommand{\vub}{|V_{ub}/V_{cb}|}
\newcommand{\vus}{|V_{us}|}

\newcommand{\beq}{\begin{equation}}
\newcommand{\eeq}{\end{equation}}

\begin{document}



\author{ {\large\bf Andrzej J.~Buras${}^{1}$ and Robert Buras${}^{2}$}\\
  \ \\
  {\small\bf ${}^{1}$ Physik Department,
    Technische Universit\"at M\"unchen,} \\
  {\small\bf D-85748 Garching, Germany} \\
  {\small\bf ${}^{2}$ Max-Planck-Institut f\"ur Physik 
 (Werner-Heisenberg-Institut),}\\ 
{\small \bf F\"ohringer Ring 6, 80805 M\"unchen, Germany}\\ 
}
  
\date{}
\title{
{\normalsize\sf
\rightline{TUM-HEP-285/00}
\rightline{MPI-TH/2000-30}
}
\bigskip
{\LARGE\bf
A Lower Bound on \boldmath{$\sin 2\beta$} from \\
Minimal Flavour Violation
}}

\maketitle
\thispagestyle{empty}

\phantom{xxx} \vspace{-9mm}

\begin{abstract}
  We point out that there exists an {\it absolute} lower bound on
  $\sin 2\beta$ in all models with minimal flavour violation (MFV), that
  do not have any new operators beyond those present in the Standard
  Model and in   which all flavour changing transitions are governed 
  by the CKM  matrix with no new phases beyond the KM phase. 
 This bound depends only on $\vcb$, $\vub$ and the hadronic
 parameters $\hat B_K$, $F_{B_d}\sqrt{\hat B_d}$ and $\xi$
 relevant for the CP-violating parameter $\varepsilon$ and the
 $B^0_{d,s}-\bar B^0_{d,s}$ mixings. Performing a simple
scanning over the present ranges of these parameters we find 
$\sin 2\beta\ge 0.34$. We illustrate  how this bound could become
stronger
when our knowledge of the parameters in question improves and
when the {\it upper} bound on the $B^0_{s}-\bar B^0_{s}$ mixing
($(\Delta M)_s$) will be experimentally known.
Provided the future accurate measurements of $\sin 2\beta$ through
the CP asymmetry in $B_d^0(\bar B_d^0)\to \psi K_S$
  will confirm the low values recently 
reported by BaBar and Belle, there is a likely possibility that
this class of models will be excluded. This would firmly imply the
necessity of new CP-violating phases and/or new effective
operators in the weak effective Hamiltonians for $K^0-\bar K^0$
and $B^0_{d,s}-\bar B^0_{d,s}$ mixings.
We also point out that within the MFV models
there exists also an absolute lower
bound on the angle $\gamma$. We find $\sin\gamma\ge 0.24$.  This
lower bound could become stronger in the future.

 \end{abstract}

\newpage
\setcounter{page}{1}
\setcounter{footnote}{0}

\section{Introduction}
\setcounter{equation}{0}
The recent measurements of the time dependent CP asymmetry
$a_{\psi K_S}$ in 
$B_d^0 (\bar B^0_d)\to\psi K_S$ decays by BaBar \cite{BaBar} and
Belle \cite{Belle} indicate that the value of the angle $\beta$
in the unitarity triangle could turn out to be substantially smaller than
expected on the basis of the standard analysis of the unitarity
triangle within the Standard Model (SM) and the CDF measurement
\cite{CDFB} of $a_{\psi K_S}$ reported last year. 
Indeed the measurements
\begin{equation}\label{sinexp}
(\sin 2\beta)_{\psi K_S} =\left\{ \begin{array}{ll}
0.12\pm0.37\pm0.09 & {\rm (BaBar)}~\cite{BaBar}, \\
0.45\pm 0.44 \pm 0.08 &{\rm (Belle)}~ \cite{Belle},\\
0.79\pm 0.42 &{\rm (CDF)}~ \cite{CDFB}
\end{array} \right.
\end{equation}
imply the grand average
\be\label{grand}
 (\sin 2\beta)_{\psi K_S}= 0.42\pm0.24~.
\label{ga}
\ee
This should be compared with the results of global analyses of the
unitarity triangle within the SM which dependent on the error estimates 
give 
\begin{equation}\label{sinth}
(\sin 2\beta)_{\rm SM} =\left\{ \begin{array}{ll}
0.75\pm0.06 & ~\cite{Stocchi}, \\
0.73\pm 0.20 &~ \cite{ALI00},\\
0.63\pm 0.12 & ~ \cite{SCHUNE},
\end{array} \right.
\end{equation}
where the last two results represent $95\%$ C.L. ranges.
Similar results can be found in \cite{AJBLH,Parodi}.
Clearly, in view of the large spread of experimental results and
large statistical errors in (\ref{sinexp}), the SM estimates in
(\ref{sinth}) are compatible with the experimental value
of $(\sin 2\beta)_{\psi K_S}$ in (\ref{grand}). Yet the small
values of $\sin 2\beta$ found by BaBar and Belle might indicate
new physics contributions to $B^0_d-\bar B^0_d$ and $K^0-\bar K^0$
mixings. In particular as discussed recently in several papers
\cite{NIR00,SW,NK00,XING} new CP violating phases in $B^0_d-\bar B^0_d$ 
mixing could be
responsible for small values of $\sin 2\beta$ in (\ref{grand}).
Indeed in this case the asymmetry $a_{\psi K_S}$ measures
$\sin 2(\beta+\theta_{\rm new})$ and choosing appropriately 
$\theta_{\rm new}$ one can obtain
 agreement with the results of BaBar and Belle.

On the other hand as stressed in \cite{NIR00} the SM estimates of
$\sin 2\beta$ are sensitive to the assumed ranges for the parameters
\begin{equation}\label{par}
\vcb,\quad \vub, \quad \hat B_K, \quad \sqrt{\hat B_d} F_{B_d}, 
\quad \xi
\end{equation}
that enter the standard analysis of the unitarity triangle.
The parameter $\xi$ is defined in (\ref{107x}).
While for ``reasonable ranges" (see table~1) of these parameters,
values of $\sin 2\beta \le 0.5$ are  excluded, such low
values  could still be possible within the SM if some of the
parameters in (\ref{par}) were chosen outside these ranges.  
In particular, for $\vub\le 0.06$ or
$\hat B_K\ge 1.3$ or $\xi\ge 1.4$,  values  lower than  0.5 for
$\sin 2\beta$ could be obtained within the SM. We agree
with these findings.

In the present letter we will assume the ``reasonable ranges"
for the parameters in (\ref{par}) as given in table~1.
The question then arises whether small values of
 $\sin 2\beta$ could still
be obtained from an analysis of the unitarity triangle in
the extentions of the SM which do not contain any new phases.
In this context we would like to point out that there exists
an {\it absolute} non-trivial lower bound for $\sin 2\beta$ in  
 models with {\it minimal flavour violation} (MFV) \cite{CDGG,UUT}, that
 do not have any new operators beyond those present in the SM and in
which all flavour changing transitions are governed by the CKM
matrix \cite{CAB} 
with no new phases beyond the KM phase. The SM, several versions of the
Minimal Supersymmetric Standard Model (MSSM) and the Two Higgs Doublet
Models I and II belong to this class. Interestingly, this absolute
lower bound on $\sin 2\beta$, which basically follows from
$\varepsilon$ and $(\Delta M)_d$, depends only on
the parameters in (\ref{par}) that are common to all  models in this 
class.  It can also be influenced significantly by the
measurement of $(\Delta M)_s$.

  Now, as pointed out recently in \cite{UUT} there exists a {\it universal}
  triangle in this class of models  that can be determined in 
  the near   future from the ratio
  $(\Delta M)_d/(\Delta M)_s$ and from $\sin 2\beta$ measured first through
  the CP asymmetry in $B_d^0\to \psi K_S$ \cite{SANDA} and later in
  $K\to\pi\nu\bar\nu$ decays \cite{BB4}.  
 Also suitable ratios of the branching
  ratios for $B\to X_{d,s}\nu\bar\nu$ and $B_{d,s}\to\mu^+\mu^-$ and
  the angle $\gamma$ measured by means of CP asymmetries in B decays
  can be used for this determination \cite{UUT}. 
    In this context, the implications of the analysis presented
 below are fourth-fold:

  \begin{itemize}
 \item
 The present ranges for the parameters in (\ref{par}) give a non-trivial
 lower bound for the angle $\beta$ in the universal unitarity
 triangle which reads
\be\label{abs}
\sin 2\beta \ge 0.34~.
\ee
\item
This bound shows that even the central value in (\ref{grand})
can be accommodated in principle within the MFV models.
No new CP violating phases are necessary.
\item
On the other hand, as illustrated below, the reduction in the
uncertainties of the parameters in (\ref{par}) and the measurement
of $(\Delta M)_s$ may result in a much stronger lower bound
on $\sin 2\beta$, that may turn out to be inconsistent with
the future improved experimental values of $(\sin 2\beta)_{\psi K_S}$.
\end{itemize}
The implications of the latter possibility would be rather profound.
With the measurement of $(\sin 2\beta)_{\psi K_S}$ alone one would be
able to conclude that one has to go beyond the concept of the MFV
and that new CP violating phases and/or new local
operators in the weak effective Hamiltonians for $K^0-\bar K^0$
and $B^0_{d,s}-\bar B^0_{d,s}$ mixings are necessary to describe
the data.

 Finally
\begin{itemize}
\item
There exists an absolute lower bound on the angle $\gamma$ which
reads
\be\label{absg}
\sin \gamma \ge 0.24~.
\ee
Also this bound could become stronger in the future.
\end{itemize}
\section{The Lower Bound on \boldmath{$\sin 2\beta$}}
In order to demonstrate that a lower bound on $\sin 2\beta$ exists in the
MFV models we use the Wolfenstein parametrization 
\cite{WO} of the CKM matrix and its generalization to include
higher order terms in $\lambda$ \cite{BLO}. Two of the Wolfenstein
parameters, $\lambda$ and $A$,
are determined from semi-leptonic K and B
decays sensitive to the elements $\vus$ and $\vcb$ respectively:
\be
\lambda=\vus=0.22, \qquad  A=\frac{\vcb}{\lambda^2}=0.826\pm0.041~.
\ee
As the decays in question are tree level decays with large
branching ratios this determination is to an excellent approximation
independent of any possible physics beyond the SM. The remaining
two parameters, $\varrho$ and $\eta$, describe the unitarity triangle.
In particular, the apex of this triangle, as shown in fig.~1, is
given by
\cite{BLO}
\begin{equation}\label{CKM4}
\bar\varrho=\varrho (1-\frac{\lambda^2}{2})~,
\qquad
\bar\eta=\eta (1-\frac{\lambda^2}{2})~.
\end{equation}
The lengths CB, CA and BA are equal respectively to 
\begin{equation}\label{2.94a}
1,\qquad
R_b \equiv  \sqrt{\bar\varrho^2 +\bar\eta^2}
= (1-\frac{\lambda^2}{2})\frac{1}{\lambda}
\left| \frac{V_{ub}}{V_{cb}} \right|,
\qquad
R_t \equiv \sqrt{(1-\bar\varrho)^2 +\bar\eta^2}
=\frac{1}{\lambda} \left| \frac{V_{td}}{V_{cb}} \right|.
\end{equation}

\begin{figure}
\begin{center}
\begin{picture}(0,0)%
\epsfig{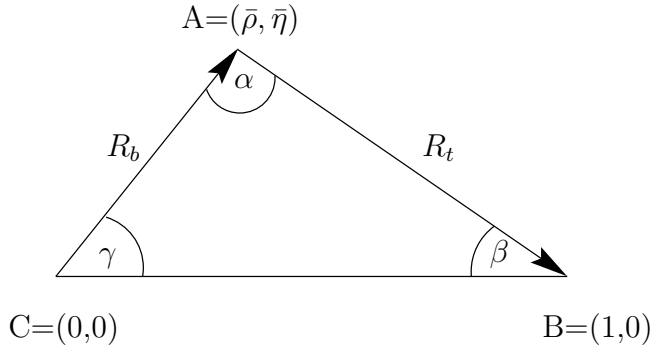}%
\end{picture}%
\setlength{\unitlength}{4144sp}%
\begingroup\makeatletter\ifx\SetFigFont\undefined%
\gdef\SetFigFont#1#2#3#4#5{%
  \reset@font\fontsize{#1}{#2pt}%
  \fontfamily{#3}\fontseries{#4}\fontshape{#5}%
  \selectfont}%
\fi\endgroup%
\begin{picture}(3342,2040)(4141,-4741)
\put(7336,-4741){\makebox(0,0)[lb]{\smash{\SetFigFont{12}{14.4}{\rmdefault}{\mddefault}{\updefault}B=(1,0)}}}
\put(4726,-3661){\makebox(0,0)[lb]{\smash{\SetFigFont{12}{14.4}{\rmdefault}{\mddefault}{\updefault}$R_b$}}}
\put(6616,-3661){\makebox(0,0)[lb]{\smash{\SetFigFont{12}{14.4}{\rmdefault}{\mddefault}{\updefault}$R_t$}}}
\put(5491,-3256){\makebox(0,0)[lb]{\smash{\SetFigFont{12}{14.4}{\rmdefault}{\mddefault}{\updefault}${\boldmath \alpha}$}}}
\put(4681,-4291){\makebox(0,0)[lb]{\smash{\SetFigFont{12}{14.4}{\rmdefault}{\mddefault}{\updefault}${\boldmath \gamma}$}}}
\put(7021,-4291){\makebox(0,0)[lb]{\smash{\SetFigFont{12}{14.4}{\rmdefault}{\mddefault}{\updefault}${\boldmath \beta}$}}}
\put(4141,-4741){\makebox(0,0)[lb]{\smash{\SetFigFont{12}{14.4}{\rmdefault}{\mddefault}{\updefault}C=(0,0)}}}
\put(5176,-2896){\makebox(0,0)[lb]{\smash{\SetFigFont{12}{14.4}{\rmdefault}{\mddefault}{\updefault}A=($\bar \rho,\bar \eta$)}}}
\end{picture}
    \end{center}
    \caption[]{Unitarity Triangle.}
    \label{fig:utriangle}
\end{figure}

Now, the experimental value for the CP violating parameter 
$\varepsilon$ combined
with the theoretical calculation of box diagrams describing 
$K^0-\bar K^0$ mixing gives the constraint for
$(\bar\varrho,\bar\eta)$ in the form of the following 
hyperbola \cite{AJBLH}:
\begin{equation}\label{100}
\bar\eta \left[(1-\bar\varrho) A^2 \eta_2 F_{tt}
+ P_c(\varepsilon) \right] A^2 \hat B_K = 0.226~.
\end{equation}
Here $\hat B_K$ is a  non-perturbative parameter
and $P_c(\varepsilon) =0.31\pm0.05$ \cite{NLOS2}  summarizes 
charm--charm and charm--top contributions in the SM.
The new physics contributions to $P_c(\varepsilon)$
are negligible within the class of the MFV models considered here 
\cite{UUT}.
Most important for our considerations is the function $F_{tt}$
that in the SM results from box diagrams with top
quark exchanges. Beyond the SM $F_{tt}$ is modified by new
particle exchanges.
$\eta_2$ is a short distance
QCD correction  related to $F_{tt}$. At NLO it
can be modified from its SM value \cite{NLOS1} by new physics
contributions to the relevant Wilson coefficients at scales
$\ord(\mw)$.  

Next, the measurement of the $B^0_d-\bar B^0_d$
mixing (the mass difference $(\Delta M)_d$) determines 
$R_t$ in the unitarity triangle of fig.~1 through 
\begin{equation}\label{RT}
R_t= 1.26~ \frac{ R_0}{A}\frac{1}{\sqrt{F_{tt}}}~,
\end{equation}
where
\be\label{R0}
 R_0= \sqrt{\frac{(\Delta M)_d}{0.47/{\rm ps}}}
          \left[\frac{200~\mev}{F_{B_d} \sqrt{\hat B_d}}\right]
          \sqrt{\frac{0.55}{\eta_B}}~.
\ee

Here $F_{tt}$  is the function present also in (\ref{100}),
$\hat B_d$ is a non-perturbative parameter analogous to $\hat B_K$, 
$F_{B_d}$ is
the $B_d$ meson decay constant and 
$\eta_B$ is the short distance QCD factor, calculated within the
SM in \cite{NLOS1,NLOS3}.

The most important feature of the formulae (\ref{100}) and
(\ref{RT}) relevant for the discussion below is that in the context
of the standard analysis of the unitarity triangle the different 
MFV models can be characterized by the value of the function $F_{tt}$.
Moreover, as explained below,
the new physics effects cancel in the ratio $\eta_2/\eta_B$.

The fact that new physics effects in $\varepsilon$
and  $B^0_{d,s}-\bar B^0_{d,s}$ mixings within the MFV models
can be described by a single function has been stressed in
particular by Ali and London \cite{ALI00}, who introduced the quantity
$f$ related to $F_{tt}$ through
\be
F_{tt}=S_0(\mt)~ (1+f)~.
\ee
Here $S_0$ is the Inami-Lim function \cite{IL} resulting within the SM from
box diagrams with top quark exchanges. In the SM $f=0.$ 
While Ali and London argued
that the universality of the quantity $f$ is approximate, we would like
to stress that in fact in each order of perturbation theory $f$ and
consequently $F_{tt}$ must be exactly the same for $\varepsilon$,
$B^0_{d}-\bar B^0_{d}$ mixing and $B^0_{s}-\bar B^0_{s}$ mixing. 
This follows from the fact that at scales $\ord(\mw)$ and higher
scales, at which new particles are integrated out, Wilson
coefficients of the relevant $\Delta F=2$ operators proportional
to $(V_{ts}^*V_{td})^2$, $(V_{tb}^*V_{td})^2$ and $(V_{tb}^*V_{ts})^2$ 
are exactly the same. The differences between these three cases
show up only in the hadronic parameters in (\ref{par}), which are not
included in $F_{tt}$ but factored out as seen in (\ref{100}) and
(\ref{RT}). Similarly the QCD corrections $\eta_2$ and $\eta_B$ 
differ only from each other by different renormalization
group evolutions below the scales $\ord(\mw)$ and consequently
the ratio $\eta_2/\eta_B$  does not
depend on new physics contributions.

In view of these remarks it will be convenient in what follows
to use the SM values  $\eta_2=0.57$ \cite{NLOS1},
$\eta_B=0.55$ \cite{NLOS1,NLOS3} and absorb all QCD corrections
related to new physics contributions into $F_{tt}$.

We are now ready to demonstrate the existence of
the lower bound on $\sin 2\beta$ in question. 
Noting that
\be\label{ss}
\sin 2\beta=\frac{2\bar\eta(1-\bar\varrho)}{R^2_t}
\ee
and combining (\ref{100}) and (\ref{RT}) one finds \cite{BLO}
\be\label{main}
\sin 2\beta=\frac{1.26}{ R^2_0\eta_2}
\left[\frac{0.226}{A^2 \hat B_K}-\bar\eta P_c(\varepsilon)\right]
\ee
whereby the first term in the parenthesis is larger
than the second term by a factor of
2--3. This dominant term is independent
of $F_{tt}$ and involves the QCD corrections only in 
the ratio $\eta_2/\eta_B$.  Consequently it is independent of $\mt$ and 
the new parameters in the extensions of the SM. The dependence on 
new physics is only present in  $\bar\eta$ entering the second term
that would be absent if the charm contribution to $\varepsilon$ was
negligible.
In particular for $\bar\varrho>0$, the value of $\bar\eta$ decreases
with increasing $F_{tt}$. 
In principle new physics could also contribute to $P_c(\varepsilon)$,
but in all known examples of MFV models such contributions are
negligible \cite{UUT}.

In spite of the sensitivity of the second term in (\ref{main}) to
new physics contributions, there exists an absolute lower bound on 
$\sin 2\beta$ in the MFV models, simply because for $\hat B_K>0$  
the unitarity of the CKM matrix implies
\be\label{RB}
0\le \bar\eta \le R_b~.
\ee 
At first sight one would think that the lower bound for $\sin 2\beta$
is attained for $\bar\eta = R_b$, but this is clearly not the 
case as $\bar\eta$ depends on the values of the parameters in (\ref{par}).
Consequently there is a correlation between the values of the
two terms in (\ref{main}).  Moreover, not arbitrary values of
$A$, $\hat B_K$, $\sqrt{\hat B_d} F_{B_d}$ and $F_{tt}$ are simultaneously
allowed by the constraints (\ref{100}) and (\ref{RT}). For instance
for given values of $A$, $R_b$ and $F_{tt}$ an approximate lower
bound for $\hat B_K$ follows from (\ref{100}) \cite{BBL}:
\be\label{bk}
\hat B_K \ge \left[ A^2 R_b(2.6 A^2 F_{tt}+1.4)\right]^{-1}~.
\ee

Thus while the discussion presented above demonstrates that
an absolute lower bound on $\sin 2\beta$ in the MFV models
exists,  $(\sin 2\beta)_{\rm min}$ for given values of the
parameters in (\ref{par}) can only be found numerically by using
the constraints (\ref{100}) and (\ref{RT}) and scanning $F_{tt}$
in the full range allowed by these constraints, the unitarity
of the CKM matrix ($1-R_b\le R_t\le 1+R_b$) and the size
of the $B^0_s-\bar B^0_s$ mixing. In the latter case $R_t$ can
be determined by measuring $(\Delta M)_s$ and using 
\begin{equation}\label{107x}
R_t= 0.82~
\xi_{\rm eff}\sqrt{\frac{(\Delta M)_d}{0.47/{\rm ps}}},
\qquad
\xi_{\rm eff}=\xi \sqrt{\frac{14.6{\rm /ps}}{(\Delta M)_s}},
\qquad
\xi = 
\frac{F_{B_s} \sqrt{\hat B_{B_s}}}{F_{B_d} \sqrt{\hat B_{B_d}}}.
\end{equation}
The existing lower bound on $(\Delta M)_s$ implies
an upper bound on $R_t$ that is much stronger than the unitarity
bound $R_t\le 1+R_b$. Using the values of the parameters in
table~\ref{tab:inputparams} we find $R_t^{\rm max}=1.03$.
This bound eliminates the solutions with $\gamma\ge 90^\circ $.
The range for $F_{tt}$ consistent with (\ref{RT}), (\ref{107x})
and $1-R_b\le R_t$ is then given by
\be\label{range}
\left[\frac{1.26  R_0}{A R_t^{\rm max}}\right]^2 \le
F_{tt}\le \left[\frac{1.26  R_0}{A(1-R_b)}\right]^2~.
\ee
In practice 
the range for $F_{tt}$ consistent also with (\ref{100}) 
 is smaller than given in (\ref{range}) as
the upper limit in (\ref{range}) corresponds to $\bar\eta=0$.

Before presenting our numerical results we would like to
recall that the weak sensitivity of $\sin 2\beta$ extracted
from $\varepsilon$ and $B^0_{d}-\bar B^0_{d}$ mixing within the
SM to the value of the top quark mass has been stressed long
time ago by Rosner \cite{Rosner}. The analytic formula
for $\sin 2\beta$ in (\ref{main}), exhibiting this feature,
has been presented in \cite{BLO}. Recently the weak dependence
of $\sin 2\beta$ on new physics contributions in a number
of SUGRA models belonging to the class of MFV models has been
emphasized by Ali and London \cite{ALI00}. 
The formula (\ref{main}), together with the range $0\le f\le 0.75$
considered in \cite{ALI00}, gives the explanation of their findings.

In the MSSM without any "SUGRA-constraints" the range 
$0\le f\le 1.13$ is still allowed \cite{EP00}.
Moreover, it is conceivable that MFV models can be constructed
for which $F_{tt}$ or $f$ are very different from those considered
in \cite{ALI00} and \cite{EP00}. In fact the  bounds
in (\ref{range}) together with the values in table~1 give
$1.3\le F_{tt}\le 15.3 $, or equivalently $-0.4\le f\le 5.5$.
In spite of this, our numerical analysis below demonstrates
that $\sin 2\beta$ exhibits a rather moderate dependence
on $F_{tt}$ in the full range allowed by (\ref{100}),
(\ref{RT}) and (\ref{range}) and that an absolute non-trivial 
lower bound on $\sin 2\beta$ exists in the MFV models.

\section{Numerical Analysis}
In our numerical analysis
we have used the standard parametrization of the CKM matrix \cite{PDG}
which is slightly more accurate than the improved Wolfenstein
parametrization of \cite{BLO}.
In table~\ref{tab:inputparams}, we list two ranges for the
relevant input parameters corresponding to the present uncertainties
and possible future reduced uncertainties. The relevant references
to table~\ref{tab:inputparams} can be found in \cite{PDG,EP99}.
The value of the running current top quark mass $\mt$ defined at 
$\mu=\mt^{Pole}$  is given here for completeness in order
to allow the translation from $F_{tt}$ to $f$.
With $\mt=165\pm 5 \gev$ one has
\be\label{ref}
F^{\rm SM}_{tt}=S_0(\mt)=2.35\pm0.11~.
\ee
We prefer to use $F_{tt}$ instead of $f$ as the small error on $\mt$
can then be taken automatically into account. We are aware of the fact
that other authors would possibly use slightly different ranges for the
input parameters. Still, table~\ref{tab:inputparams} is representative 
for the present situation.

Finally, although $\vub$ does not enter the formulae
(\ref{100}), (\ref{RT}) and (\ref{main}) explicitly, its value enters
our analysis in the same manner as in the standard analysis of the unitarity
triangle \cite{AJBLH}. See for instance (\ref{bk}) and (\ref{range}).

\begin{table}[thb]
\caption[]{The ranges of the input parameters.
\label{tab:inputparams}}
\vspace{0.4cm}
\begin{center}
\begin{tabular}{|c|c|c|c|}
\hline
{\bf Quantity} & {\bf Central} & {\bf Present} & {\bf Future} \\
\hline
$|V_{cb}|$ & 0.040 & $\pm 0.002$ & $\pm 0.001$     \\
$\vub$ & $0.090$ & $\pm 0.018 $ &
$\pm 0.009 $  \\
$\hat B_K$ & 0.85 & $\pm 0.15$ & $\pm 0.07$  \\
$\sqrt{\hat B_d} F_{B_{d}}$ & $200\mev$ & $\pm 40\mev$ & 
$\pm 20\mev$ \\
$\mt$ & $165\gev$ & $\pm 5\gev$ & $\pm 2\gev$  \\
$(\Delta M)_d$ & $0.471/\mbox{ps}$ & $\pm 0.016/\mbox{ps}$ 
& $\pm 0.008~\mbox{ps}^{-1}$  \\
$(\Delta M)_s$ & & $>14.6/\mbox{ps}$ & $>16.6/\mbox{ps}$ 
\\
$\xi$ & $1.16$ & $\pm 0.07$ & $\pm 0.04$ \\
$\mc(\mc)$ & $1.30\gev$    & $\pm 0.05\gev$ & $\pm 0.05\gev$ \\
\hline
\end{tabular}
\end{center}
\end{table}

In fig.~2 we show $(\sin 2\beta)_{\rm min}$ as a function of $F_{tt}$
for the two choices of uncertainties in the input parameters 
given in table~\ref{tab:inputparams}. To this end we have scanned
independently all the input parameters within the ranges of this
table. We observe that the dependence of $(\sin 2\beta)_{\rm min}$
on $F_{tt}$ is rather weak. For $F_{tt}\ge 13.5~(7.8)$ in the case
of the present (future) scanning there are no solutions for
$\sin 2\beta$ in the ranges of the parameters considered.
The absolut lower
bound for $\sin 2\beta$ is found to be
\begin{equation}\label{absolut}
(\sin 2\beta)_{\rm min} =\left\{ \begin{array}{ll}
0.34 & {\rm Present}~, \\
0.48 & {\rm Future}~.\\
\end{array} \right.
\end{equation}
We would like to emphasize that these bounds should be considered
as conservative. After all they have been obtained by scanning 
independently
all parameters in question. Had we used the error analysis of 
\cite{Stocchi}, the bounds in (\ref{absolut}) would have been 
considerably stronger.

\begin{figure}
\unitlength1mm
\begin{picture}(121,85)
  \put(25,5){\psfig{file=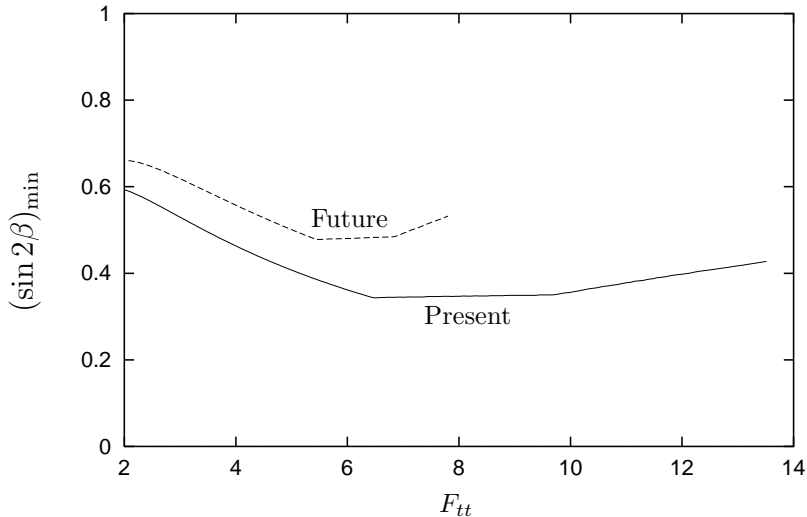,width=10.4cm}} 
  \put(60,40){\footnotesize Future}
  \put(75,27){\footnotesize Present}
  \put(77,2){\footnotesize $F_{tt}$}

\put(23,38){%
\makebox(0,0)[b]{\shortstack{$(\sin 2 \beta)_{\rm min}$}}%
}%

\end{picture}
  \caption{Lower bound for $\sin 2\beta$ as a function of
 $F_{tt}$ for present and future ranges of the input parameters.}
\label{lowerbound}
\end{figure}

From fig.~2 one can extract the lower bounds for particular MFV models
by setting $F_{tt}$ to the appropriate values. A number of supersymmetric
MFV models has been reviewed by Ali and London \cite{ALI00},
where references to the original literature can be found. 
Using the results of \cite{ALI00}
we find as characteristic values  
$F_{tt}=3.0$, $F_{tt}=3.4$, $F_{tt}=4.3$ for  minimal SUGRA models,
non-minimal SUGRA models and non-SUGRA models with EDM constraints
respectively.
Setting in addition $F_{tt}=2.46$ and $F_{tt}=5.2$ for the SM
and the MSSM version of \cite{EP00} respectively, we obtain
\begin{equation}\label{models}
(\sin 2\beta)_{\rm min} =\left\{ \begin{array}{ll}
0.57~ (0.64) & {\rm SM}~, \\
0.53~ (0.62) & {\rm MSUGRA}~,\\
0.50~ (0.59) & {\rm NMSUGRA}~,\\
0.44~ (0.54) & {\rm NSUGRA}~,\\
0.40~ (0.49) & {\rm MSSM}~.\\
\end{array} \right.
\end{equation}
The numbers in parentheses correspond to the future ranges of
the input parameters. Our results for the models analyzed in \cite{ALI00}
are compatible with the ranges for $\sin 2\beta$ presented there.
We observe that the present $(\sin 2\beta)_{\rm min}$ in the NSUGRA 
models and in the MSSM are
in the ball park of the grand average in (\ref{grand}).

The anatomy of the bound is given in table~\ref{ANA}, where we show
$(\sin 2\beta)_{\rm min}$ as a function of $\hat B_K$ and $\vcb$
with all the remaining parameters scanned within the present
and future ranges. 
There is no solution for the set of parameters corresponding to the
last entry in table~\ref{ANA}.
The numbers in parentheses are discussed
below. 
On the basis of this table and additional numerical analysis
we find the following features in accordance with
(\ref{main}):

\begin{itemize}
\item
$(\sin 2\beta)_{\rm min}$ decreases with increasing $\hat B_K$
and $\vcb$ and decreasing $\vub$ and $F_{B_d}\sqrt{\hat B_d}$.
\item
In the ranges considered, the dependence of $(\sin 2\beta)_{\rm min}$
on $\hat B_K$, $\vcb$ and $F_{B_d}\sqrt{\hat B_d}$ is 
stronger than on $\vub$. This is evident from (\ref{main}) in
which $\vub$ is not explicitly present but affects the bound only 
through the value of $\bar\eta$ in the sub-leading term and 
 through its impact on the allowed ranges of the
remaining parameters.
\item
One can check that the dependence of $(\sin 2\beta)_{\rm min}$
on $\hat B_K$ and $A$ can be
approximately given by a single variable $\tau=A^2 \hat B_K$. This
is clear from (\ref{main}). The observed small departure
from this regularity is caused by the correlations between various
parameters as discussed below equation (\ref{RB}).
\end{itemize}

\begin{table}[thb]
\caption[]{ Values of $(\sin 2\beta)_{\rm min}$ in the
MFV models for specific values of $\hat B_K$ and $\vcb$ 
with  remaining parameters in the ``present" and "future"
ranges. The  values in the parentheses show the impact of the
measurement of $(\Delta M)_s$ with $\xi_{\rm eff}\ge 1.0$.
\label{ANA}}
\begin{center}
\begin{tabular}{|c|c|c|c|c||c|}\hline
 $\hat B_K $& $ \vcb $ & Present &
  Future \\ \hline
       & $0.038$ & 0.59~(0.62)  & 0.70~(0.70) \\
 $0.70$& $0.040$ & 0.54~(0.58)  & 0.65~(0.65) \\
       & $0.042$ & 0.49~(0.55)  & 0.59~(0.61)  \\
 \hline
      & $0.038$ & 0.49~(0.62)  &  0.60~(0.68) \\
 $0.85$& $0.040$ & 0.44~(0.58)  & 0.54~(0.65) \\
       & $0.042$ & 0.40~(0.55)  & 0.49~(0.61)  \\
 \hline
      & $0.038$ & 0.42~(0.62)  &  0.51~(0.68) \\
 $1.00$& $0.040$ & 0.38~(0.58)   & 0.46~(0.64) \\
       & $0.042$ & 0.34~(0.55)  & 0.42~(---)  \\
 \hline
 \end{tabular}
\end{center}
\end{table}
Next we would like to emphasize that the measurement of $B^0_s-\bar B^0_s$
mixing may significantly improve the lower bound on $\sin 2\beta$
considered here. Measuring $R_t$ by means of (\ref{107x})
could provide a lower bound on $R_t$ in addition to the
known upper bound. This in turn
would exclude high values of $F_{tt}$ as seen in (\ref{RT}). The numbers
in the parentheses in table~\ref{ANA} show 
the impact of the $(\Delta M)_s$ measurement with
 $\xi_{\rm eff}\ge 1.0$, or equivalently $(\Delta M)_s\le 19.6/{\rm ps}$ for
$\xi=1.16$. We observe that the impact of the measurement of 
$(\Delta M)_s$ is very large, in particular in the case of the
present scenario. The values $\sin 2\beta\le 0.55$ are excluded.
With increasing  $\xi_{\rm eff}$ the impact becomes stronger. It is
weaker for smaller $\xi_{\rm eff}$, but even for $\xi_{\rm eff}=0.90$,
$\sin 2\beta\le 0.41$ and
$\sin 2\beta\le 0.51$ are excluded in the present and future scenario
respectively.

We note that the previously found dependence of $(\sin 2\beta)_{\rm min}$
on $\hat B_K$ and $\vcb$ is strongly affected by the lower bound
on $R_t$. In particular the dependence on $\hat B_K$ is negligible.
This is related to the fact that with $\xi_{\rm eff}\ge1.0$, $R_t$ is
confined to $0.82\le R_t \le 1.03 $ and $(\sin 2\beta)_{\rm min}$ is
governed by the values of $R_t$ and $R_b$. With decreasing 
$\xi_{\rm eff}$ the $\hat B_K$ dependence becomes again visible.

One remark on  the $\xi_{\rm eff}$ dependence is in order.
As we have seen $(\sin 2\beta)_{\rm min}$ increases with increasing
$\xi_{\rm eff}$. This feature is valid only for $\gamma$ in the
first quadrant. For $\gamma$ in the second quadrant
$(\sin 2\beta)_{\rm min}$ decreases with increasing
$\xi_{\rm eff}$, as already noticed in \cite{NIR00}.

Finally we would like to point out that the absolute lower bound
on $\sin 2\beta$ implies within the MFV models an absolute lower
bound on the angle $\gamma$. We find
\begin{equation}\label{absgam}
(\sin \gamma)_{\rm min} =\left\{ \begin{array}{ll}
0.24 & {\rm Present}~, \\
0.39 & {\rm Future}~,\\
\end{array} \right.
\end{equation}
with $\gamma$ in the first quadrant. The second quadrant in the
MFV models is excluded through the lower bound on $(\Delta M)_s$.

In analogy to (\ref{models}) we find 
\begin{equation}\label{gmodels}
(\sin \gamma)_{\rm min} =\left\{ \begin{array}{ll}
0.68~ (0.82) & {\rm SM}~, \\
0.58~ (0.70) & {\rm MSUGRA}~,\\
0.53~ (0.63) & {\rm NMSUGRA}~,\\
0.46~ (0.54) & {\rm NSUGRA}~,\\
0.40~ (0.47) & {\rm MSSM}~.\\
\end{array} \right.
\end{equation}

The anatomy of $(\sin \gamma)_{\rm min}$ is given in table~\ref{ANAG}.
As in the case of $(\sin 2 \beta)_{\rm min}$ the impact of the
measurement of $(\Delta M)_s$ is very significant.

\begin{table}[thb]
\caption[]{ Values of $(\sin \gamma)_{\rm min}$ in the
MFV models for specific values of $\hat B_K$ and $\vcb$ 
with  remaining parameters in the ``present" and "future"
ranges. The  values in the parentheses show the impact of the
measurement of $(\Delta M)_s$ with $\xi_{\rm eff}\ge 1.0$.
\label{ANAG}}
\begin{center}
\begin{tabular}{|c|c|c|c|c||c|}\hline
 $\hat B_K $& $ \vcb $ & Present &
  Future \\ \hline
       & $0.038$ & 0.40~(0.77)  & 0.62~(0.80) \\
 $0.70$& $0.040$ & 0.38~(0.76)  & 0.58~(0.78) \\
       & $0.042$ & 0.37~(0.74)  & 0.54~(0.77)  \\
 \hline
      & $0.038$ & 0.31~(0.77)  &  0.45~(0.79) \\
 $0.85$& $0.040$ & 0.30~(0.76)  & 0.43~(0.78) \\
       & $0.042$ & 0.29~(0.74)  & 0.42~(0.77)  \\
 \hline
      & $0.038$ & 0.25~(0.77)  &  0.37~(0.79) \\
 $1.00$& $0.040$ & 0.25~(0.76)   & 0.35~(0.78) \\
       & $0.042$ & 0.24~(0.74)  & 0.34~(---)  \\
 \hline
 \end{tabular}
\end{center}
\end{table}

\section{Conclusions}
 
  We have pointed out that there exists an {\it absolute} lower bound on
  $\sin 2\beta$ in the MFV  models, that
  do not have any new operators beyond those present in the Standard
  Model and in   which all flavour changing transitions are governed 
  by the CKM  matrix with no new phases beyond the KM phase. 
 This bound depends only on $\vcb$, $\vub$ and the non-perturbative
 parameters $\hat B_K$, $F_{B_d}\sqrt{\hat B_d}$ and $\xi$
 relevant for the CP-violating parameter $\varepsilon$ and the
 $B^0_{d,s}-\bar B^0_{d,s}$ mixings.
 The present ranges of these parameters imply $\sin 2\beta\ge0.34$.

We have illustrated how the lower bound on
$\sin 2\beta$ could  become stronger
when our knowledge of the input parameters in question improves and
when the {\it upper bound} on 
$(\Delta M)_s$ will be experimentally known.
In particular, if the upper bounds on $\hat B_K$ and $\vcb$ and
lower bounds on $\vub$, $F_{B_d}\sqrt{\hat B_d}$ and $\xi_{\rm eff}$
in (\ref{107x}) will be improved,
 $(\sin 2\beta)_{\rm min}$ will be shifted above 0.5. 
Consequently,
provided the future accurate measurements of $a_{\psi K_S}$
will confirm the low values reported by BaBar and Belle,  
there is a likely possibility that
all MFV models will be excluded. This would firmly imply the
necessity of new CP-violating phases and/or new effective
operators in the weak effective Hamiltonians for $K^0-\bar K^0$
and $B^0_{d,s}-\bar B^0_{d,s}$ mixings. 

We have also pointed out that the lower bound
on $\sin 2\beta$ implies within the MFV models an absolute lower
bound on the angle $\gamma$. This provides an additional test
of the MFV models once precise measurements of $\gamma$
will be available.

Clearly other measurements, in particular those of the
rare decay branching ratios and various CP asymmetries, 
will have an additional
impact on the analysis presented here, but this is
a different story. For a very recent review see 
\cite{NIR08}.

It will be exciting to watch the experimental progress in
the values of $a_{\psi K_S}$ and $(\Delta M)_s$ and the
theoretical progress on $\hat B_K$, $\vub$, $F_{B_d}\sqrt{\hat B_d}$
and $\xi$. Possibly we will know already next summer that new
CP violating phases and/or new operators in the effective
weak Hamiltonians are mandatory.

We would like to thank Martin Gorbahn for useful discussions and
critical comments on the manuscript.
This work has been supported in part by the German Bundesministerium 
f\"ur Bildung and Forschung under the contract 05HT9WOA0 and by the
Deutsche Forschungsgemeinschaft under grant No.~SFB 375.

\vfill\eject


\begin{thebibliography}{99}
\bibitem{BaBar}
D. Hitlin, BaBar collaboration, plenary talk at ICHEP 
(Osaka, Japan, July 31, 2000), SLAC-PUB-8540.
\bibitem{Belle}
H. Aihara, Belle collaboration, plenary talk at ICHEP 
(Osaka, Japan, July 31, 2000).
\bibitem{CDFB}
T. Affolder et al., CDF collaboration,
{ Phys. Rev.} {\bf D61} (2000) 072005.
\bibitem{Stocchi}
F. Caravaglios, F. Parodi, P. Roudeau, and A. Stocchi, hep-ph/0002171.
\bibitem{ALI00}
A. Ali and D. London, Eur. Phys. J. {\bf C9} (1999) 687;
hep-ph/0002167.
\bibitem{SCHUNE}
S. Plaszczynski and M.-H. Schune, hep-ph/9911280.
\bibitem{AJBLH}
A.J. Buras and R. Fleischer, hep-ph/9704376;
A.J. Buras, hep-ph/9806471, hep-ph/9905437.
\bibitem{Parodi}
S. Mele, {Phys.\ Rev.} {\bf D59} (1999) 113011;
M. Bargiotti et al., La Rivista del Nuovo Cimento, Vol. 23, N.3 (2000) 1;
 S. Schael, Phys. Rept.
{\bf 313} (1999) 293; M.~Ciuchini, E.~Franco, L.~Giusti, V.~Lubicz and 
G.~Martinelli, Nucl.\ Phys.\  {\bf B573} (2000) 201.
\bibitem{NIR00}
G. Eyal, Y. Nir, and G. Perez, hep-ph/0008009.
\bibitem{SW}
J.P. Silva and L. Wolfenstein, hep-ph/0008004.
\bibitem{NK00}
A.L. Kagan and M. Neubert, hep-ph/0007360.
\bibitem{XING}
Z.Z Xing, hep-ph/0008018.
\bibitem{CDGG}
M. Ciuchini, G. Degrassi, P. Gambino and G.F. Giudice,
{ Nucl. Phys.} {\bf B 534} (1998) 3.
\bibitem{UUT}
A.J. Buras, P. Gambino, M. Gorbahn, S. J\"ager and L. Silvestrini,
hep-ph/0007085.
\bibitem{CAB}
{ {N. Cabibbo}}, { Phys. Rev. Lett.} {\bf 10} (1963) 531;
{ M. Kobayashi and K. Maskawa},
 { Prog. Theor. Phys.} {\bf 49} (1973) 652.
\bibitem{SANDA}
 I.I.Y. Bigi and A.I. Sanda,
 Nucl. Phys. {\bf B193} (1981) 85.
\bibitem{BB4}
{ G. Buchalla and A.J. Buras}, 
{ Phys. Lett.} {\bf B333} (1994) 221; 
{ Phys. Rev.} {\bf D54} (1996) 6782.
\bibitem{WO}
{ L. Wolfenstein}, { Phys. Rev. Lett.} {\bf 51} (1983) 1945.
\bibitem{BLO}
{ A.J. Buras, M.E. Lautenbacher and G. Ostermaier,}
{ Phys. Rev.} {\bf D50} (1994) 3433.
\bibitem{NLOS2}
{ S. Herrlich and U. Nierste,}
{ Nucl. Phys.} {\bf B419} (1994) 292, 
{ Phys. Rev.} {\bf D52} (1995) 6505, 
{ Nucl. Phys.} {\bf B476} (1996) 27.
\bibitem{NLOS1}
{ A.J. Buras, M. Jamin, and P.H. Weisz,}
{ Nucl. Phys.} {\bf B347} (1990) 491.
\bibitem{NLOS3}
J. Urban, F. Krauss, U. Jentschura and G. Soff, 
{ Nucl. Phys.} {\bf B 523} (1998) 40.
\bibitem{IL}
T. Inami and C.S. Lim, Progr. Theor. Phys. {\bf 65} (1981) 297.
\bibitem{BBL}
{ G. Buchalla, A.J. Buras and M. Lautenbacher,} 
{ Rev. Mod. Phys} {\bf 68} (1996) 1125.
\bibitem{Rosner}
J.L. Rosner, in ``B Decays", ed. S. Stone, World Scientific,
Singapour (1994), page 470.
\bibitem{EP00}
A.J. Buras, P. Gambino, M. Gorbahn, S. J\"ager and L. Silvestrini,
hep-ph/0007313.
\bibitem{PDG}
{ Particle Data Group,} { Euro. Phys. J.} {\bf C 3} (1998) 1.
\bibitem{EP99}
S. Bosch, A.J. Buras, M. Gorbahn, S. J\"ager, M. Jamin,
M.E. Lautenbacher and L. Silvestrini,
{ Nucl. Phys.} {\bf B 565} (2000) 3.
\bibitem{NIR08}
Y. Nir, hep-ph/0008226.
\end{thebibliography}
\end{document}